\documentclass[prl,twocolumn,showpacs]{revtex4-1}
\usepackage{graphicx,multirow,bm}
\usepackage{amsmath}
\renewcommand{\vec}[1]{\bm{#1}}
\renewcommand{\Im}{\mathrm{Im}}

\newcommand{\ket}[1]{|\, {#1} \,\rangle}

\begin{document}
\title{Feshbach Resonances in Cesium at Ultra-low Static Magnetic Fields}
\author{D. J. Papoular$^{1,2}$, S. Bize$^3$,  A. Clairon$^3$, H. Marion$^3$, S. J. Kokkelmans$^4$, G.V. Shlyapnikov$^{1,5}$}
\affiliation{$^1$\mbox{Laboratoire de Physique Th{\'e}orique et Mod{\`e}les Statistiques, CNRS, Universit{\'e} Paris-Sud, F-91405, Orsay, France}}
\affiliation{$^2$\mbox{INO--BEC Center and Dipartimento di Fisica, Universit{\`a} di Trento, 38123 Povo, Italy}}
\affiliation{$^3$\mbox{LNE-SYRTE, Observatoire de Paris, CNRS, UPMC, F-75014 Paris, France}}
\affiliation{$^4$\mbox{Eindhoven University of Technology, P.O. Box 513, 5600 MB Eindhoven, The Netherlands}}
\affiliation{$^5$\mbox{Van der Waals--Zeeman Institute, University of Amsterdam, Science Park 904, 1098 XH Amsterdam, The Netherlands}}

\date{\today}

\begin{abstract}
We have observed Feshbach resonances for ${}^{133}\mathrm{Cs}$ atoms
in two different hyperfine states  at ultra--low static magnetic fields 
by using an atomic fountain clock.
The extreme sensitivity of our setup allows for high signal--to--noise--ratio observations
at densities of only $2\times 10^{7}\,\mathrm{cm}^{-3}$. We have reproduced these resonances using coupled--channels calculations 
which are in excellent agreement with our measurements.
We justify that these
are $s$--wave resonances involving weakly--bound states of the triplet molecular Hamiltonian, identify the resonant closed channels,
and explain the observed  multi--peak structure.
We also describe a model which precisely accounts for the collisional processes in the fountain and which explains the asymmetric shape of the observed Feshbach resonances in the regime where the kinetic energy dominates over the coupling strength.
\end{abstract}


\maketitle
The achievement of Bose--Einstein condensation
\cite{anderson:science_1995,*bradley:PRL_1995,*davis:PRL_1995} has stimulated remarkable developments in atomic physics. Ultracold atoms have found applications in metrology \cite{Guena2012} and high--precision measurements of physical constants \cite{biraben:EPJst_2009}; they can be cooled down to quantum degeneracy and used to simulate
condensed--matter systems \cite{bloch:RMP_2008,giorgini:RMP_2008}.
A fundamental feature of ultracold atomic gases, underlying most of their present applications, is that the interparticle interactions 
can be tailored at will, using scattering resonances that occur in low--energy collisions between two atoms \cite{chin:RMP_2010}.
These Feshbach resonances are usually obtained using an external static magnetic field \cite{inouye:Nature_1998}. Their accurate characterization is intimately linked to a detailed knowledge of the interatomic interaction \cite{chin:PRA_2004} and involves coupled--channels calculations \cite{verhaar_PRA2009}.

We report on the measurement of multiple Feshbach resonances
in ${}^{133}\mathrm{Cs}$ using an atomic fountain clock,
and present their theoretical characterization using the coupled--channels method.
The extreme accuracy of frequency measurements in modern atomic clocks provides the means to reveal effects of atomic collisions in a regime of very weak interactions. The excellent agreement between experimental measurements and theory confirms that the interaction between Cs atoms is now well understood and modeled.
The resonances that we analyze are unusual for two main reasons.
First, they occur at magnetic fields of the order of a few milliGauss,
which makes them the lowest--static--field resonances investigated up to now.
In these ultralow magnetic fields, the quasi--degeneracy of all collisional channels with a triplet two--atom electronic spin plays a key role and conveys a multi--peak structure to the resonances. Second, we have measured them in a regime where the kinetic energy dominates over the resonance width. In this regime, they appear in the magnetic field dependence of the clock shift
as asymmetric features  which occur close to the zero--temperature resonant
 field. 

The further experimental characterization of these low-field resonances
using density--independent interferometry \cite{hart:Nature_2007}, combined
with the enhanced sensitivity to the values of fundamental constants near a
Feshbach resonance \cite{chin:PRL_2006,borschevsky:PRA_2011}, could be used to probe the
constancy of the proton--to--electron mass ratio and the fine structure constant.
Furthermore, these resonances involve atoms in two different spin states and thus pave the way towards the study of quantum magnetism in ultracold Cesium gases containing two different hyperfine states.

\begin{figure}[h]
  \begin{minipage}{.35\columnwidth}
  \includegraphics[width=.6\linewidth]{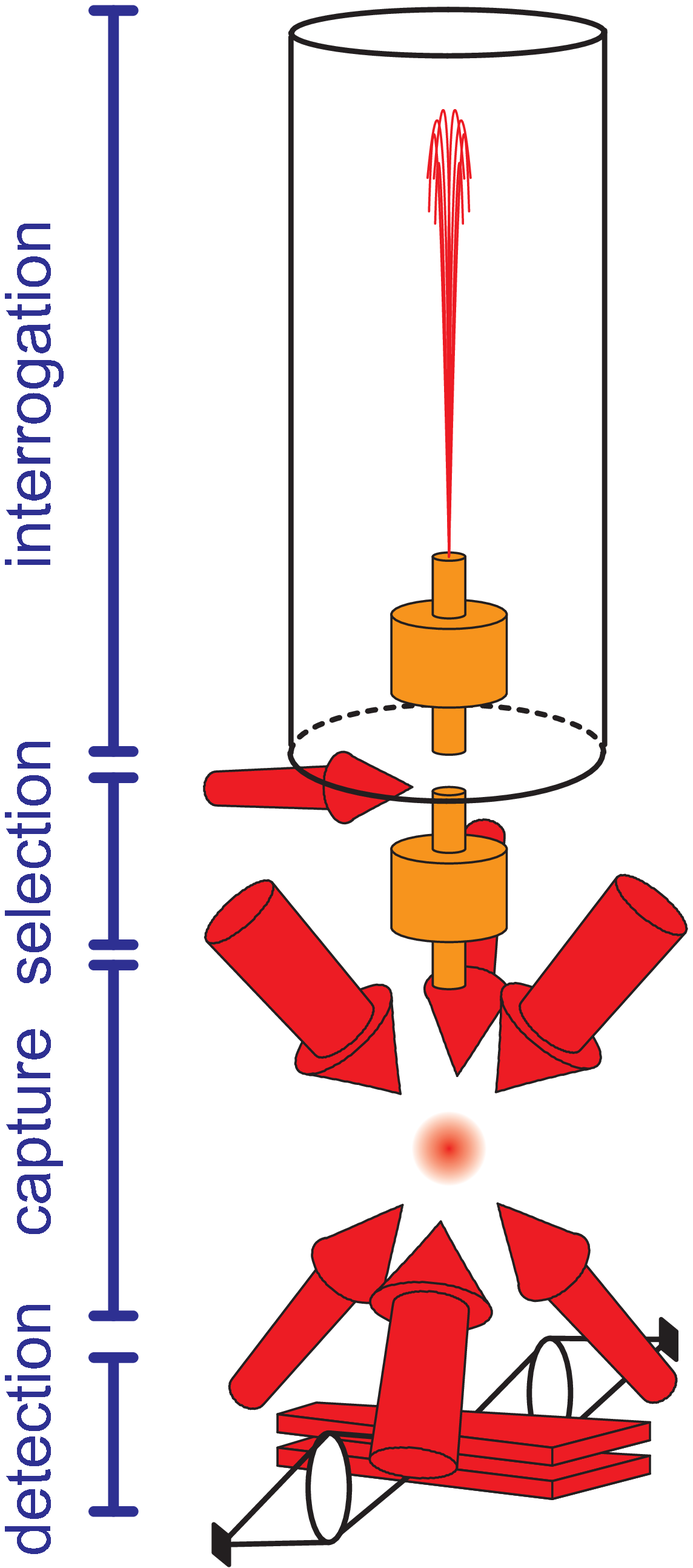}
  \end{minipage}
  \begin{minipage}{.63\columnwidth}
    \includegraphics[width=\linewidth]{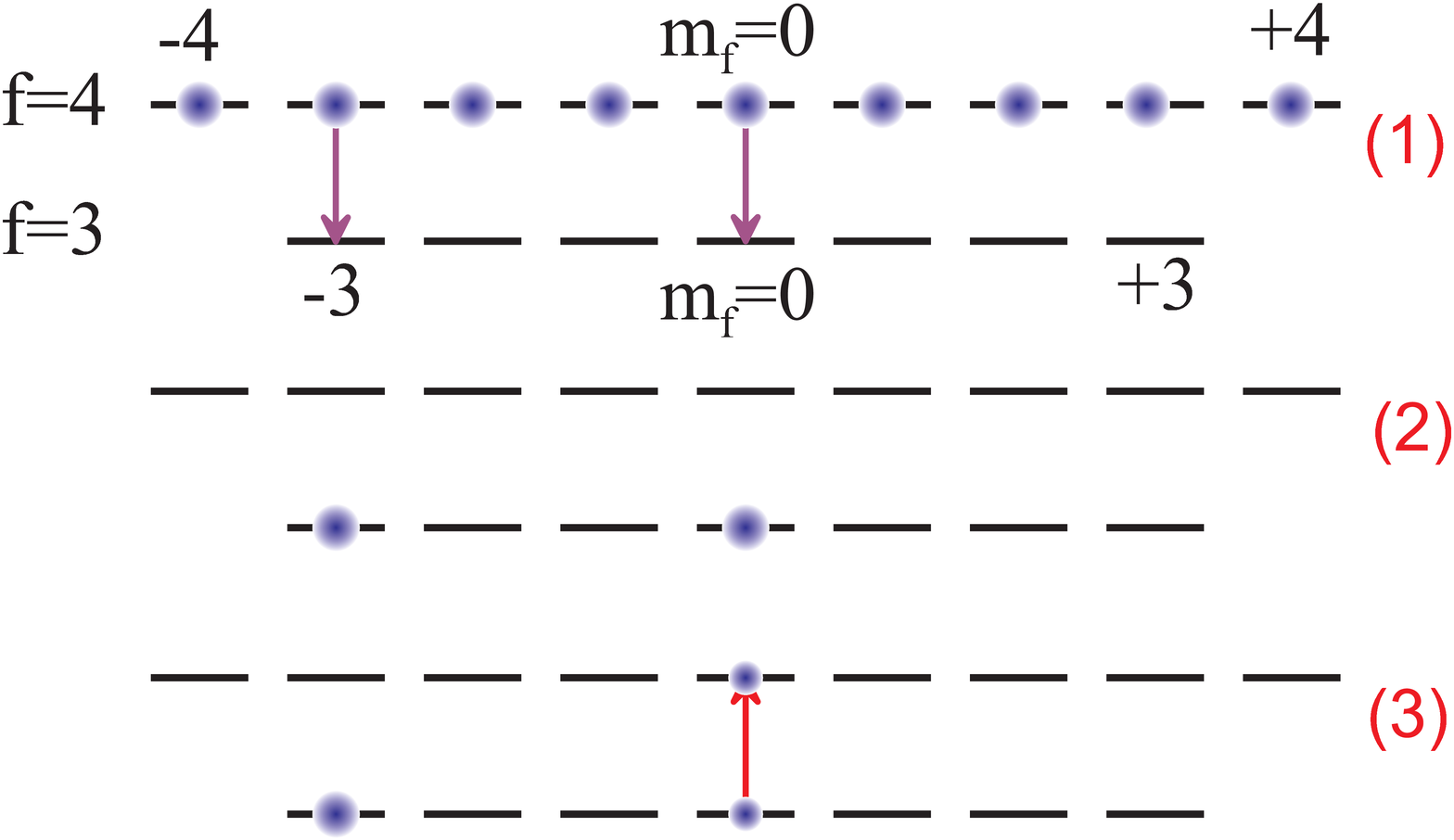}
  \end{minipage}
  \caption{\label{fig:fountain}
    Left: schematics of the atomic fountain. Right: hyperfine levels used in the experiment. Populations after launch and transitions excited for the state selection (1). A horizontal laser beam then pushes away $f=4$ atoms. Populations at the start (2) and during (3) the Ramsey interrogation.}
\end{figure}
\emph{Experimental setup.}--- The experiment is done in a fountain geometry which has already between described extensively (see e.g. \cite{Guena2012}) and which is sketched in Fig.~\ref{fig:fountain}. The frequency of the $|f=3,m_f=0\rangle \longrightarrow|f=4, m_f=0\rangle$ hyperfine transition is probed during the ballistic flight of a cloud of $^{133}$Cs atoms laser--cooled to $\sim 1$~$\mu$K. The Ramsey interrogation occurs between the upward and downward traversals of a microwave cavity. 
After the Ramsey interrogation, the atom numbers in the $f=3$ and $f=4$ hyperfine states are measured by  laser--induced fluorescence detection in order to determine the transition probability. Before the interrogation, state selection is applied to the up-going cloud by means of microwave and laser interactions (see Fig.~\ref{fig:fountain}). For the present experiments, we select not only the $|3,m_f=0\rangle$ clock state but also an additional $|3,m_f\neq 0\rangle$ state, and measure the clock frequency shifts due to this state.

\emph{Clock shift measurements of Feshbach resonances.}--- Collision--induced frequency shifts depend on elementary collisional properties but also on the atomic spatial and velocity distributions. This latter dependence is even stronger in the present experiment, first because of the evolution of the atomic cloud during the Ramsey interrogation (see e.g. \cite{Sortais2000}), and second because we are in a regime of strong sensitivity of the measured shifts to the collision energy. We determine collision shift ratios in a way that minimizes the impact of atomic distributions which are 
difficult to control with high precision. We perform interleaved frequency measurements with 3 configurations, leading to 3 measured frequencies: $\nu_{0}^{(1)}$, $\nu_{0}^{(1/2)}$ and $\nu_{0;m_f}^{(1)}$. Firstly, the $|3,m_f=0\rangle$ state is selected with the maximum possible atom number $N_0$. Secondly, the $|3,m_f=0\rangle$ state is selected with the atom number $N_0/2$. Thirdly, $N_0$ atoms in $|3,m_f=0\rangle$ are selected together with $N_{m_f}$ atoms in another chosen $|3,m_f\rangle$ state, as illustrated in Fig. \ref{fig:fountain} for $m_f=-3$. 
The expanding atomic cloud is truncated during the Ramsey cavity traversals, so that the detected atoms are only a fraction ($\sim 20\%$) of the initially selected atoms. We choose to characterize the number of atoms using the detected atoms, hence $N_0$, $N_0/2$ and $N_{m_f}$ will refer to the atom numbers as measured in the detection. For $N_0$ and $N_0/2$, this is the sum of $|3,0\rangle$ and $|4,0\rangle$ atoms since some atoms which are initially in 
the state $|3,0\rangle$ are excited to the state $|4,0\rangle$ during the Ramsey interrogation.
A crucial feature of our experiment is to perform the microwave excitation for state selection  with the (interrupted--)adiabatic passage method described in \cite{Pereira2002} in order to ensure quasi--identical space and velocity distributions for all states ($m_f=0$ and $m_f\neq0$) and all configurations, notably the first and the second one.
We can prepare the third configuration with any of the six
$m_f=\pm 1,\pm2,\pm3$ states.

\begin{figure*}[ht]
  \includegraphics[width=.32\textwidth]
  {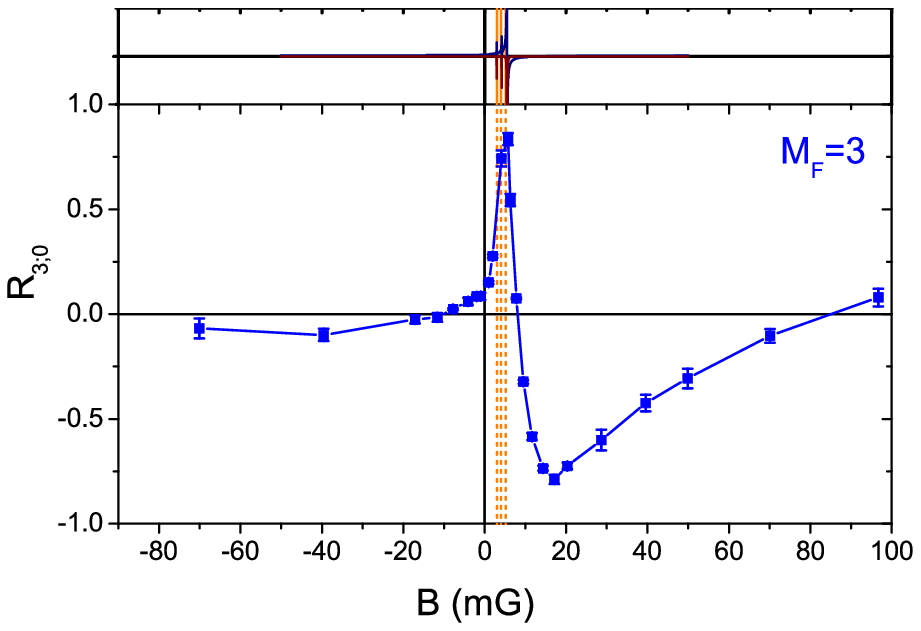}
  \includegraphics[width=.32\textwidth]
  {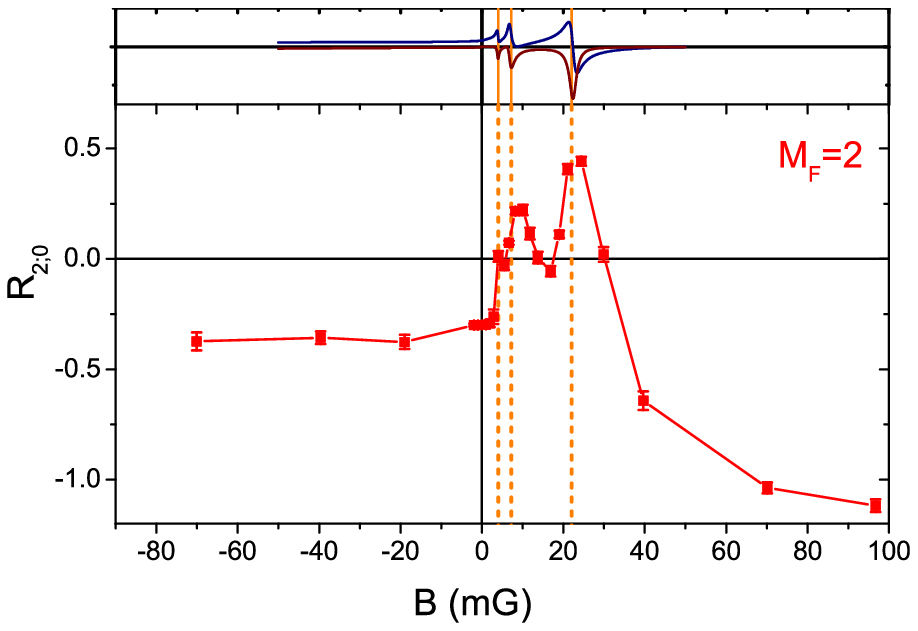}
  \includegraphics[width=.32\textwidth]
  {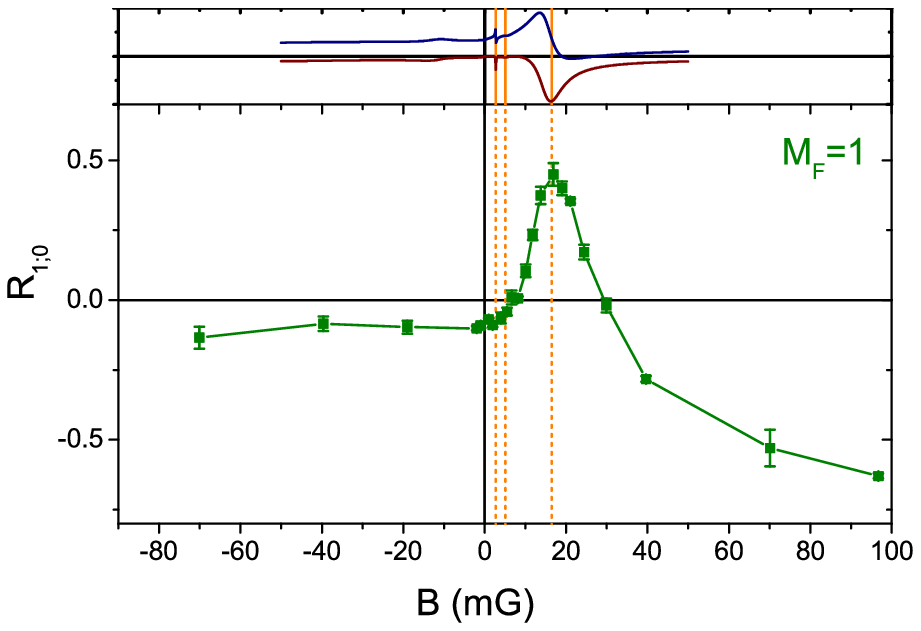}
  \caption{\label{fig:statres_syrte}
    Feshbach resonances at low static magnetic fields. Left:
    $M_F=3$; middle: $M_F=2$; right: $M_F=1$.
    Lower graphs: measured collision frequency shift ratio $R_{m_f;0}$
derived from the experiment as a function of the static magnetic field.
Upper graphs: theoretical values for the scattering length.
The vertical dashed lines show the theoretical resonant field values.
Experiment (effective temperature $\sim 900\,\mathrm{nK}$)
and theory ($T=0$) are completely independent (see text).}
\end{figure*}

In a given configuration, the frequency shift of the $|3,0\rangle \longrightarrow|4,0\rangle$ transition  is given by:
\begin{equation}
  \delta\nu=
  n_0 \rho_0 K_0(B, \mathcal{D}_{r,v}) +
  n_{m_f} \rho_{m_f} K_{m_f}(B, \mathcal{D}_{r,v})
  \ ,
\end{equation}
where $n_0$ and $n_{m_f}$ are the detected atom numbers,
$\rho_{0}$ and $\rho_{m_f}$ are the effective densities per detected atom,
and
$K_0$ and $K_{m_f}$ are the collision shifts scaled to the effective densities. These functions include collisional properties,
which depend on the magnetic field $B$. They also depend
on the space and velocity distributions $\mathcal{D}_{r,v}$, and more generally on the fountain geometry.
Starting from the measured frequency shifts and the detected atom numbers, we compute the shift per detected $m_f=0$ atom, $A_{0;0}$,
and the additional shift $A_{m_f;0}$
due to the $m_f\neq 0$ population, 
per detected $m_f\neq 0$ atom:
\begin{equation}
    \begin{split}
  A_{0;0}&=\frac{\nu_{0}^{(1)}-\nu_{0}^{(1/2)}}{N_0-N_0/2}=\rho_0 K_0(B, \mathcal{D}_{r,v})\\
  A_{m_f; 0}&=\frac{\nu_{0;m_f}^{(1)}-\nu_{0}^{(1)}}{N_{m_f}}=\rho_{m_f} K_{m_f}(B, \mathcal{D}_{r,v})
    \end{split}
\end{equation}
Our state selection ensures quasi-identical distributions for all states, so that $\rho_0 \simeq \rho_{m_f}$, and, hence,
$A_{m_f; 0}/A_{0;0}\simeq K_{m_f}(B, \mathcal{D}_{r,v})/K_0(B, \mathcal{D}_{r,v})$. This quantity is as close to intrinsic collisional properties as possible in our experiment. Notably, it does not depend on the detected atom numbers  $N_0$ and $N_{m_f}$. 
Typically, $N_0\sim 5\cdot 10^{6}$
and $N_{m_f}\approx N_0$. The corresponding effective density during the Ramsey interrogation, $N_0 \rho_0 \sim 2\cdot 10^{7}\,\mathrm{cm}^{-3}$, is many orders of magnitudes lower than in typical quantum gas experiments. The mean free path is $\sim 35$~m and the mean time between collisions is $\sim 5000\,\mathrm{s}$, i.e. 3 orders of magnitude longer than the experimental cycle.

We have determined $A_{0;0}$, $A_{m_f; 0}$ and $R_{m_f;0}=A_{m_f; 0}/A_{0;0}$ for all $m_f$ states as a function of the magnetic field $B$
from $0$ to $100\,\mathrm{mG}$. The magnetic field is known via the spectroscopy of the first--order--sensitive
$|3,m_f=1\rangle \rightarrow|4,m_f=1\rangle$ transition.
It is stable to $\sim 40\,\mathrm{nG}$ and homogeneous to better than $10^{-2}$. Our measurements of $R_{m_f;0}(B)$ 
are shown in
Fig.~\ref{fig:statres_syrte}. 
The magnetic field keeps the same downward orientation over the entire height of the fountain
to avoid Majorana transitions and to ensure a good control of the quantization axis.
Under these conditions, 
selecting a $-m_f$ state for a measurement is equivalent to 
probing the $+m_f$ state
with the field $-B$. Therefore, we plot measurement results with $B<0$ 
which are, in fact, taken with a negative $m_f$ state.
For all 3 states, we observe a dramatic dependence of $R_{m_f;0}$  on $B$. Instead, we measure no significant change of the clock collision shift $A_{0;0}(B)$, at a level limited by the dependence of this quantity on the effective density $\rho_0$, which itself is quite sensitive to variations in the atomic distributions $\mathcal{D}_{r,v}$. Within these limits, 
$K_0(B, \mathcal{D}_{r,v})$ remains constant over the entire range of our experiments. It is equal to the large negative clock shift which 
affects Cs fountain clocks \cite{Gibble1993,Clairon1995,Leo2001}. 
Hence, the observed behavior of $R_{m_f;0}(B)$ relates to
$K_{m_f}(B, \mathcal{D}_{r,v})$, which we attribute to Feshbach resonances either in the $|3,0;3,m_f\rangle$ or the $|4,0;3,m_f\rangle$ channel.

The precise control of the magnetic field and the high signal--to--noise ratio of the data allow for a stringent comparison to two theoretical approaches:
\textit{(i)} a coupled--channels
calculation of the scattering length characterizing 
interactions at zero temperature as a function of $B$, and
\textit{(ii)} a finite--temperature model of the
clock collision shift in the fountain geometry,
which explains the asymmetric shape
 of the observed resonances.

\begin{figure*}[ht]
  \includegraphics[angle=-90,width=.32\textwidth]
  {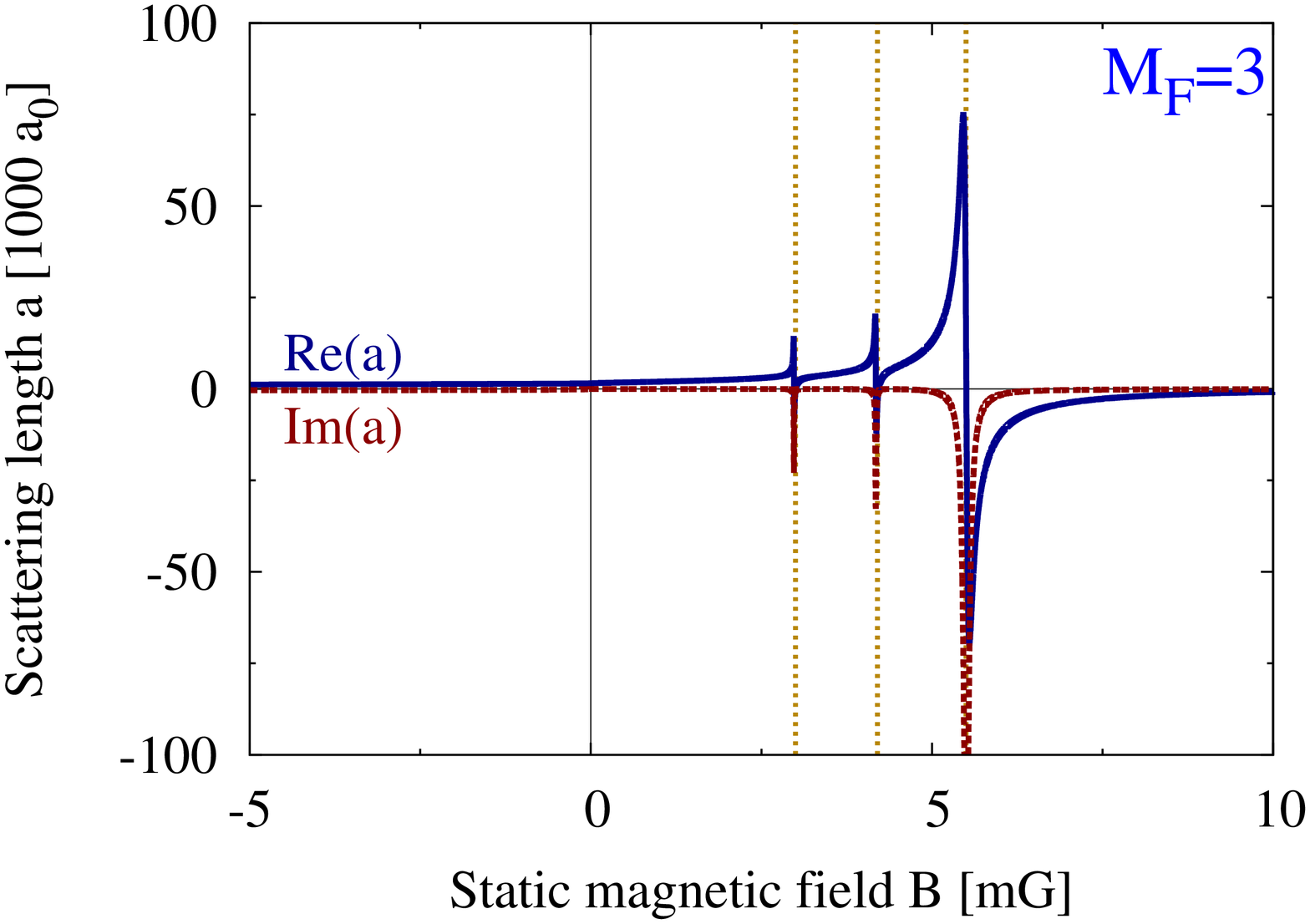}
  \includegraphics[angle=-90,width=.32\textwidth]
  {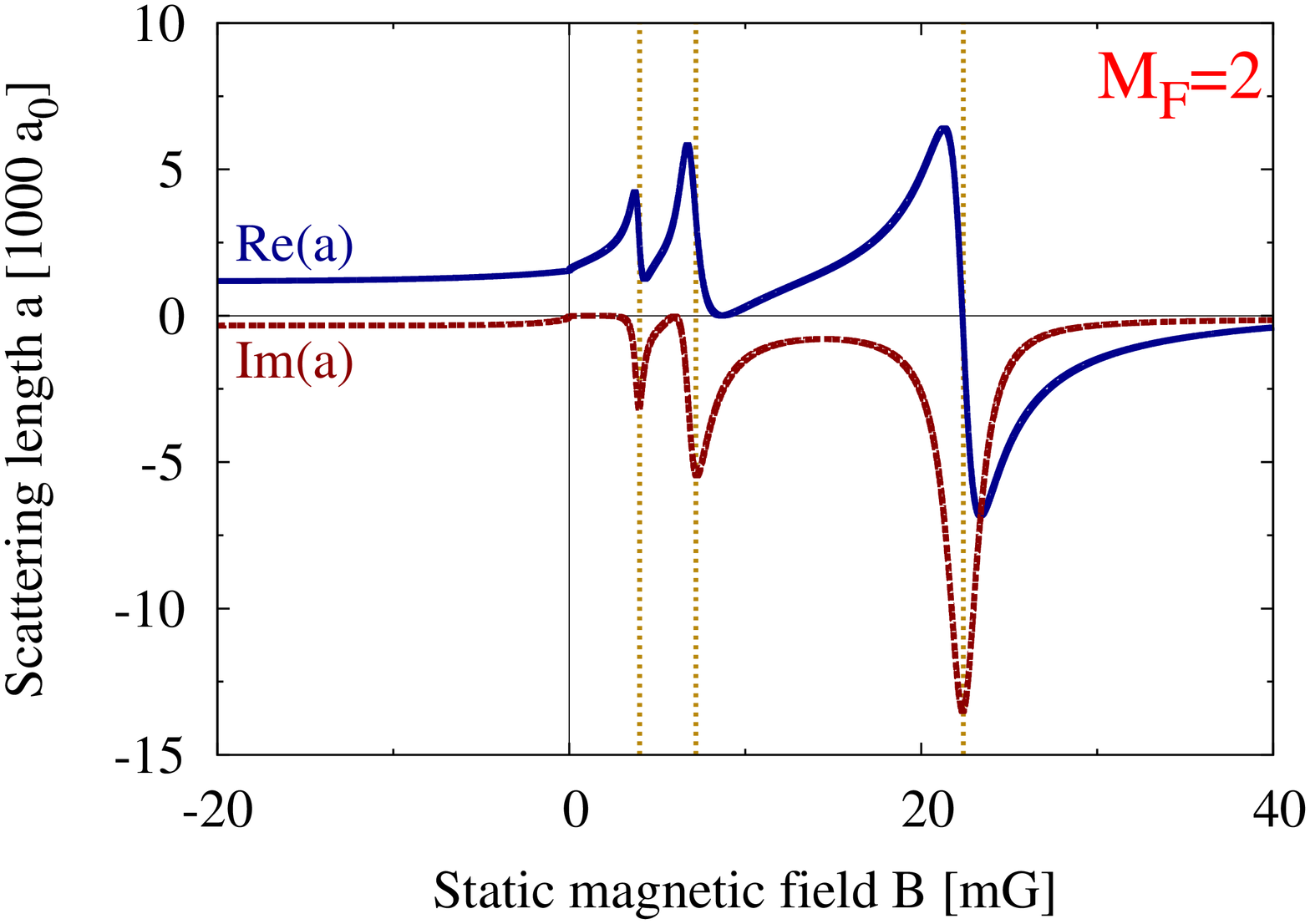}
  \includegraphics[angle=-90,width=.32\textwidth]
  {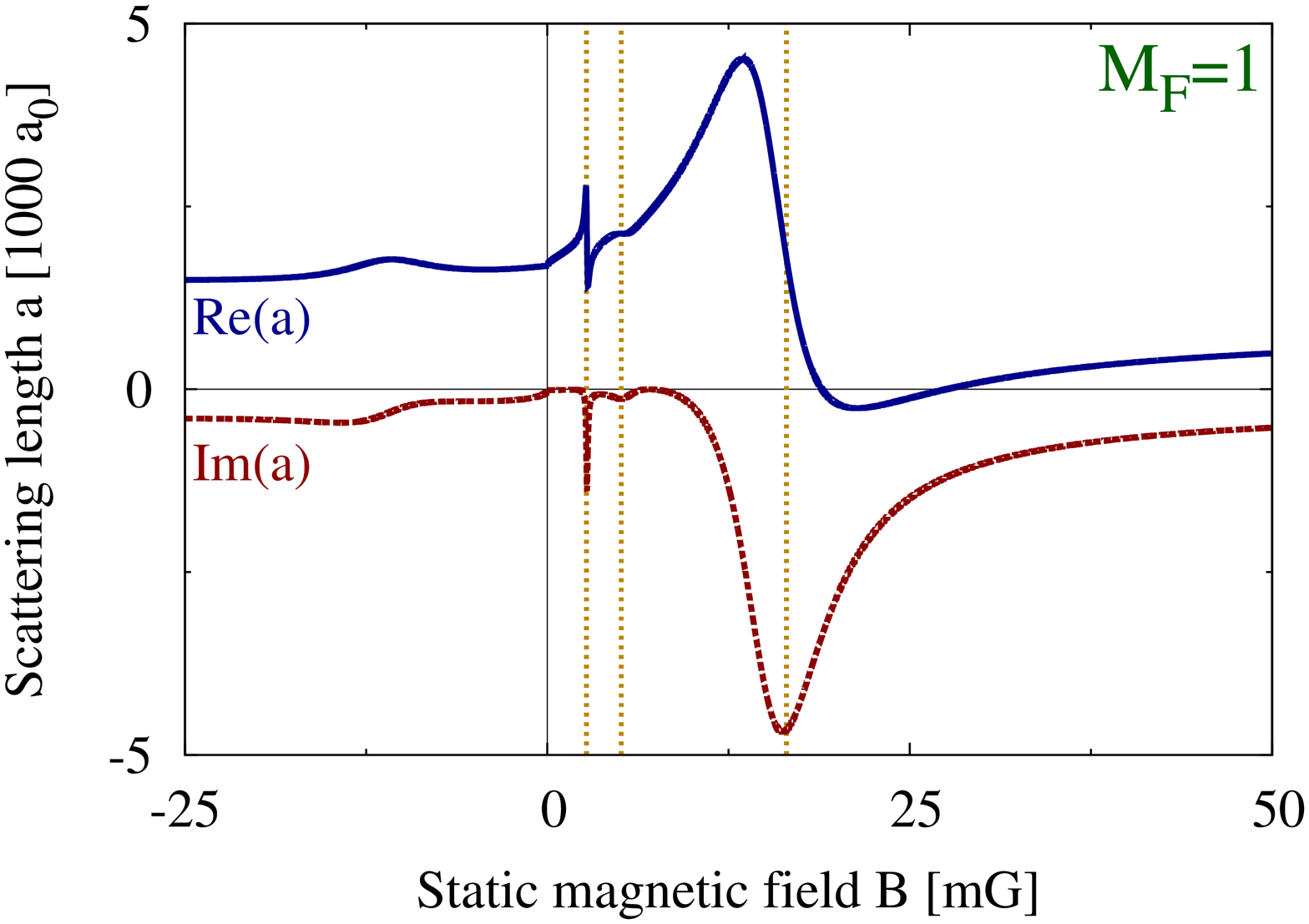}
  \caption{\label{fig:statres_fpfm}
    Zoom--in onto the upper graphs 
    in Fig.~\ref{fig:statres_syrte}.
    Left: $M_F=3$; middle: $M_F=2$; right: $M_F=1$.}
\end{figure*}

\emph{Calculation of the scattering length.}--- We describe the system in the center--of--mass frame of the atom pair.
Neglecting the spin--spin interaction, which yields no significant contribution to our observables,
the interaction is spatially isotropic. We limit our analysis to $s$--wave interactions governed by the following Hamiltonian
\cite{verhaar_PRA2009}:
\begin{equation}
  H=\frac{p^2}{2\mu}+V_\mathrm{el}(r)+V_\mathrm{hf}+V_\mathrm{Z}\ ,
\end{equation}
where $r$ is the interatomic distance, $p$ is its conjugate momentum, and $\mu=m/2$ is the reduced mass of the atom pair. The central part of the interaction is given by $V_\mathrm{el}(r)=V_S(r)P_S+V_T(r)P_T$,
where $P_S$ and $P_T$ are the projectors onto the electronic--singlet and triplet subspaces. The term
$V_\mathrm{hf}=
  a_\mathrm{hf}
  (\vec{s}_1\cdot\vec{i}_1+\vec{s}_2\cdot\vec{i}_2)
  /\hbar^2$
is the hyperfine interaction, where $\vec{s}_j$ and $\vec{i}_j$ are the spin operators of the electron and the nucleus of atom $j$. The operator
$V_\mathrm{Z}=2\mu_B B\, S_z$ is the
Zeeman term \footnote{The small coupling of the magnetic field to the nuclear spins is included in our numerical calculations and does not affect their results.},
with $\mu_B$ being the Bohr magneton and $S_z=s_{1z}+s_{2z}$ being the total electronic spin projection along the quantization axis $\vec{e}_z$.

We calculate the magnetic field dependence of the scattering
length associated with the zero--energy scattering state
corresponding to the levels populated in the experiment.
The Hamiltonian $H$ conserves the projection $M_F$ of
the total two--atom spin $\vec{F}=\vec{f}_1+\vec{f}_2$,
where $\vec{f}_j=\vec{s}_j+\vec{i}_j$ is the total spin of atom $j$.
Therefore, this scattering state has a definite value of the total spin projection $M_F$, on which the scattering length $a_{M_F}(B)$ depends.
For large interatomic separations, the atoms are in the Zeeman--dressed state related to the (Bose--symmetrized) two--atom state $\ket{f_1=4,m_1=0;f_2=3,m_2=M_F}$, where the quantum numbers $f_j$ and $m_j$ define the magnitude and projection of the total spin $\vec{f}_j$.
The experimental results shown in Fig.~\ref{fig:statres_syrte} correspond to $M_F=3$, $2$, and $1$, respectively.

The scattering state $\ket{\vec{\Psi}_{M_F,B}}$
has $10$ coupled components if $M_F=3$, and $13$ and $14$
components for $M_F=2$ and $M_F=1$, respectively.
We evaluate it numerically using the coupled--channels approach \cite{verhaar_PRA2009}, our implementation of which is described in \cite{papoular_PHD2011}. 
The accumulated--phase boundary condition \cite{verhaar_PRA2009}
is applied at $r_0=20\,\mathrm{a_0}$, and the asymptotic behaviour of the zero--energy scattering state is enforced at
$r_\mathrm{max}=1000\,\mathrm{a_0}$.
All resulting differential systems are solved using Stoermer's rule with adaptive stepsize control \cite{nr3_CUP2007}.
The values used for the accumulated--phase parameters, the hyperfine interaction constant $a_\mathrm{hf}$, and the electronic
potentials $V_S$ and $V_T$
are the same as those used in \cite{papoular_PRA2010}.

\begin{table}[h]
  \begin{center}
    \begin{tabular}{|c|c||c|c||c|c|}
      \hline
      \multicolumn{6}{|c|}{Resonance positions $B_\mathrm{res}$ [mG]}\\
      \hline
      \multicolumn{2}{|c||}{$M_F=1$} &
      \multicolumn{2}{c||}{$M_F=2$} &
      \multicolumn{2}{c|}{$M_F=3$}  \\
      meas. & calc. & meas. & calc. & meas. & calc. \\
      \hline
                 & $ 2.7$     &       & $4.0$ &           & $3.0$\\
                 & $ 5.1$     & $8$   & $7.2$ &           & $4.2$\\
      $18 \pm 3$ & $16.5$     & $25$  & $22$  & $5 \pm 1$ & $5.5$\\
      \hline
    \end{tabular}
    \caption{\label{tab:statrespos}
      Measured and calculated resonance positions.}
  \end{center}
\end{table}
Our results for the $s$--wave scattering length $a_{M_F}(B)$
are shown in Fig.~\ref{fig:statres_fpfm}, for $M_F=3$, $2$, and $1$.
The occurrence of inelastic processes
(such as the decay towards the lower--energy states having
$f_1=f_2=3$) causes $a$ to have a non--vanishing imaginary part \cite{landau_QM1991} and the resonances appear as smooth dispersive features (rather than as the divergences of the lossless case).
The calculated resonance positions, corresponding to the minima of $\Im(a)$, are shown in Table~\ref{tab:statrespos}. The calculated positions for the broadest resonances compare favorably to those determined from the experimental clock--shift measurements (Fig.~\ref{fig:statres_syrte}).
The predicted multiple--peak structure
is clearly visible in the experimental data for $M_F=2$.

Our numerical analysis includes only $s$--wave interactions,
and the fact that it recovers the measured resonance positions proves that these are $s$--wave resonances.
The triplet potential $V_T$ supports a very weakly bound state,
with the binding energy
$|E_T|=\hbar^2/(2\mu a_T^2)\approx
  h\cdot 5\,\mathrm{kHz}=
  \mu_B\cdot 4\,\mathrm{mG}$,
where
$a_T=2400\,a_0$ is the scattering length associated with $V_T$ \cite{chin:PRA_2004}.
For a given value of $M_F$, the 
two--atom internal states $\ket{f_1=4,f_2=3,F,M_F}$ are electronic--triplet for all allowed odd values of $F$.
For $B=0$, each of these triplet channels supports the weakly--bound triplet state,
yielding $N^T_M$ degenerate bound states (energy $-|E_T|$), where $N^T_M$ is the number of triplet channels with the quantum number $M_F$.  For non--zero, albeit small, magnetic fields, the coupling due to $V_Z$ lifts this degeneracy,
and these $N^T_M$ states cross the threshold for different values of $B$,
causing multiple resonances.
For $M_F=3$ or $2$, there are $N^T_M=3$ triplet channels ($F=7$, $5$, or $3$), which correspond to the three predicted resonances in these two cases. For
$M_F=1$, there are $N^T_1=4$ triplet states ($F=7$, $5$, $3$, $1$); however, our coupled--channels results only show three resonances, probably because the fourth one is too narrow to be resolved in the presence of the three other peaks.
This 
multiple--resonance physics only occurs for small $B$: indeed, for values of $B$ larger than a few $|E_T|/\mu_B$, the Zeeman term $V_Z$ causes the bare weakly--bound triplet states to dissolve into the continuum.

\emph{Feshbach resonances in a fountain geometry.}--- To clarify the impact of
finite temperatures, the atomic spatial and velocity distributions $\mathcal{D}_{r,v}$, and the fountain geometry, we have
evaluated the clock shift using a simple model for  the $S$--matrix elements
$S_{\alpha\gamma}(k)$ and $S_{\beta\gamma}(k)$
describing the interaction between the clock states, $\alpha=|3,0\rangle$ and $\beta=|4,0\rangle$, and the additional state $\gamma=|3,m_f\rangle$.
The elementary clock shift due to $\gamma$ is :
\begin{equation}
    \frac{\delta\omega_{\beta\alpha}}{2\pi}=
    \frac{\hbar \rho_{\gamma}}{m k} \,
    \Im\left\{
      S_{\alpha\gamma}(k)S^{\dag}_{\beta\gamma}(k)-1
    \right\}
    \ ,
    \label{eq:clockshift}
\end{equation}
with $\rho_{\gamma}$ being the local density, both in time and space,
of atoms in the state $\gamma$, and $k=p/\hbar$ being the wavevector
for the relative motion of the two colliding atoms.
For a resonance occurring in the $\beta\gamma$ channel, we take $S_{\alpha\gamma}(k)=1$ and assume that $S_{\beta\gamma}(k)$
is given by:
\begin{equation}
    S_{\beta\gamma}(k)=1-\frac{i \Gamma_e}{E-\nu(B) + \frac{1}{2}i \Gamma_e}
    \ ,
    \label{eq:Smatrix}
\end{equation}
where $E=\hbar^2 k^2/2\mu$ is the relative kinetic energy of the colliding pair, $\nu(B)$ is the energy detuning from the resonance, and $\Gamma_e=k C_e $ is the elastic width of the resonance \cite{Moerdijk1995}, the coupling strength
$C_e$ being constant.
We have omitted the inelastic contribution to the width, $i\Gamma_i/2$, in the denominator of Eq.~(\ref{eq:Smatrix}), as our coupled--channels results imply that $\Gamma_i/C_e\ll k$. 
\begin{figure}[h]
  \includegraphics[width=0.491\columnwidth]{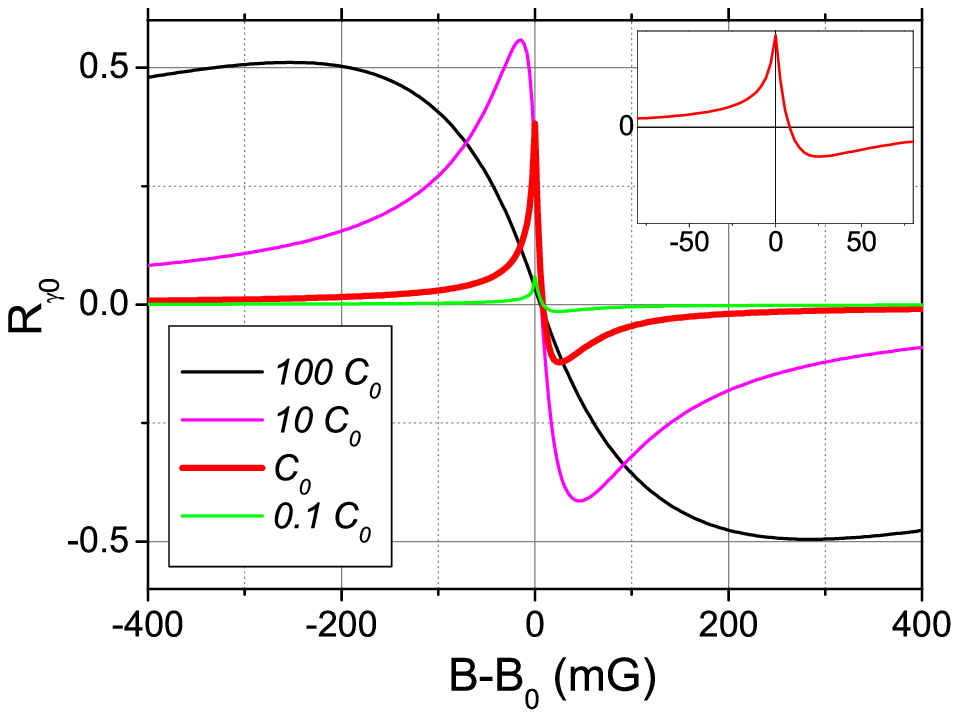}
  \quad
  \includegraphics[width=0.46\columnwidth]{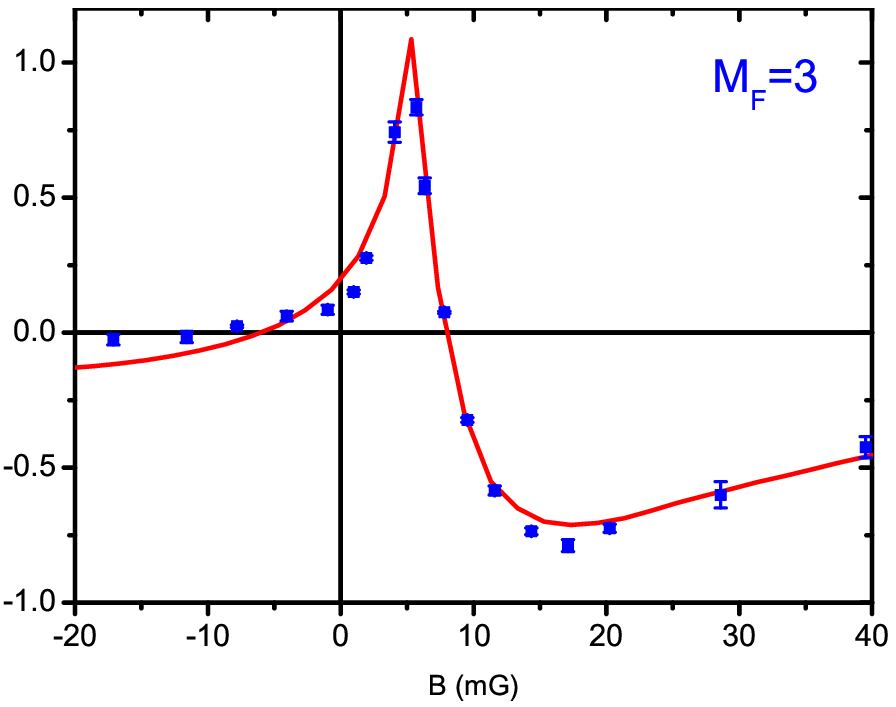}
  \caption{\label{fig:FeshbachModel}
  Left: clock shift as a function of $B$, 
numerically calculated
using Eqs.~\ref{eq:clockshift} and \ref{eq:Smatrix},
for various coupling strengths $C_e$. The value
$C_e=C_0=4\, E_\mathrm{rec}/k_\mathrm{rec}$ (red)
is close to the experimental situation;
results for $C_e=100\, C_0$ (black), $C_e=10\, C_0$ (magenta),
and $C_e=0.1\, C_0$ (green) are also shown.
Inset: close--up of the red curve for small values of $B$.
Right: fit of our model (Eqs.~\ref{eq:clockshift} and \ref{eq:Smatrix})
to the $M_F=3$ measurements.
  }
\end{figure}
The total clock shift is obtained by averaging Eq.~(\ref{eq:clockshift}) over the atomic space and velocity distribution $\mathcal{D}_{r,v}$ measured in the experiment. We calculate it using a Monte--Carlo simulation
accounting for the collisional energy distribution (corresponding to the
effective temperature $\sim 900\,\mathrm{nK}$), the decrease of the atomic density with time, and the truncation of the atomic cloud in the microwave resonator. 
Figure~\ref{fig:FeshbachModel} (left) shows the total clock shift as a function of $B$ for various coupling strengths $C_e$.
As an example, we take $\nu=2\mu_B (B-B_0)$, corresponding to a scattering length $a(B)=-C_e/(4\mu_B(B-B_0))$.
The black curve is for $C_e=400 E_\mathrm{rec}/k_\mathrm{rec}$,
where $\hbar k_\mathrm{rec}=h/\lambda$
and $E_\mathrm{rec}=\hbar^2 k_\mathrm{rec}^2 /2 m$ are the
recoil momentum and energy, and $\lambda = 852\,\mathrm{nm}$
is the laser cooling wavelength.
In this strong--coupling regime,
the resonance has a symmetrical dispersive-like shape. At any given field,
all atoms within the distribution $\mathcal{D}_{r,v}$ contribute to it,
and the collision shift reaches the unitarity limit.
The green curve ($C_e=0.4 E_\mathrm{rec}/k_\mathrm{rec}$) illustrates the weak--coupling regime, in which the 
kinetic energy exceeds the elastic width.
In this regime, the resonance curve is strongly asymmetric \footnote{This asymmetry is not due to the $B$--dependence of the background contribution
to the scattering amplitude, which is difficult to resolve in fountain--clock measurements.}.
For $B<B_0$, the resonant channel is closed and the behavior is similar to the far--detuned strong--coupling case. For $B>B_0$, the resonant channel is open. At a given field, only a fraction of the distribution $\mathcal{D}_{r,v}$ contributes significantly to the frequency shift because of the narrow elastic width. Consequently, the total clock shift is smaller than the unitarity limit value.
The experimental value $C_e=C_0=4 E_\mathrm{rec}/k_\mathrm{rec}$
(thick red) is close to the weak--coupling regime. The resonant behavior of the clock shift is clearly visible, and the inset shows that it occurs at the zero--temperature resonant field $B_0$, where
this model predicts a singularity even at finite temperature.
A fit of our model to the measurements for $M_F=3$ (Fig.~\ref{fig:FeshbachModel} right) captures the main features of the data, and in particular its asymmetry. This fit yields $B_0=5\pm 1$~mG.
Were the resonance occurring in the $\alpha\gamma$ channel, the sign of the clock shift would be reversed. Therefore this analysis, independent of our coupled--channels results, confirms that the resonance occurs in the $\beta\gamma$ channel.

We have measured multiple Feshbach resonances in ${}^{133}\mathrm{Cs}$ at ultralow magnetic fields using a fountain clock, and characterized 
them
theoretically using the coupled--channels approach. We have identified the resonant bound 
state to be the
weakly--bound state of the triplet potential
and explained their multi--peak structure. 
They have been observed in a regime where the kinetic energy dominates over the resonance width, which causes them to appear as asymmetric features in the $B$--dependence of the clock shift, as captured by
our  finite--temperature Monte--Carlo simulations.
The resonant triplet state can also be brought to resonance using a weak microwave field tuned far away from the single--atom resonance \cite{papoular_PRA2010}, thus leaving the single--atom Physics 
unaffected, which could also be useful for metrological applications.

We acknowledge many fruitful discussions with J. Dalibard, P. Rosenbusch, and C. Salomon.

%

\end{document}